\documentclass[12pt]{article}
\PassOptionsToPackage{hyphens}{url}
\usepackage[pagebackref=true,colorlinks]{hyperref}
\usepackage{amsmath,amssymb,amsthm,mathtools,amsrefs,mathrsfs,marvosym}
\usepackage[utf8]{inputenc}
\usepackage{pifont}% http://ctan.org/pkg/pifont
\usepackage{mathpazo}
\usepackage{authblk} % add possibly `sc` and `osf` options
\usepackage{bbm}
\usepackage{tikz}
\usepackage{quantikz}
\usepackage{comment}
\usepackage{fancyhdr}
\usepackage{soul}
\usepackage{dsfont}%this is for mathbb{1} but you write mathds{1}
\usepackage[normalem]{ulem}
\usepackage{longtable}
\usepackage[bottom]{footmisc}%this forces footnotes to appear on the bottom of page
\usepackage[font=small,labelfont=bf]{caption}
\usepackage{subcaption}
\usepackage{enumerate}
\usepackage[all]{xy} %This is to make higher categorical stuff
\hypersetup{
	colorlinks=true,%set true if you want colored links
    linktoc=all,%set to all if you want both sections and subsections linked
    linkcolor=violet,  %choose some color if you want links to stand out
    citecolor=violet,
    }
\usepackage{tocbasic}
\DeclareTOCStyleEntry[
  beforeskip=0.3em plus 1pt,% default is 1em plus 1pt
  pagenumberformat=\textbf
]{tocline}{section}
\makeatletter
\newcommand*{\shifttext}[2]{%
  \settowidth{\@tempdima}{#2}%
  \makebox[\@tempdima]{\hspace*{#1}#2}%
}
\makeatother
\makeatletter
\renewcommand*\env@matrix[1][\arraystretch]{%
  \edef\arraystretch{#1}%
  \hskip -\arraycolsep
  \let\@ifnextchar\new@ifnextchar
  \array{*\c@MaxMatrixCols c}}
\makeatother

\tikzset{
    gateT/.style={
        draw,
        not gate US,
        inner sep=5pt
    },
    gateO/.style={
        draw,
        circle,
        minimum width=1.67em,
        inner sep=2pt,
        append after command={
            \pgfextra {
                \fill (\tikzlastnode);
            }
        }
    }
}

\DeclareExpandableDocumentCommand{\gateT}{O{}{m}}{|[gateT,#1]| {#2} \qw}
\DeclareExpandableDocumentCommand{\gateO}{O{}{m}}{|[gateO,#1]| {#2} \qw}

\theoremstyle{plain}
\newtheorem{theorem}[equation]{Theorem}
\newtheorem{lemma}[equation]{Lemma}
\newtheorem{proposition}[equation]{Proposition}
\newtheorem{corollary}[equation]{Corollary}
\theoremstyle{definition}
\newtheorem{definition}[equation]{Definition}
\newtheorem{construction}[equation]{Construction}
\newtheorem{question}[equation]{Question}

\newtheorem{problem}[equation]{Problem}
\newtheorem{example}[equation]{Example}
\newtheorem{exercise}[equation]{Exercise}
\newtheorem*{answer}{Answer}
\newtheorem*{solution}{Solution}
\newtheorem{remark}[equation]{Remark}

\newtheorem{notation}[equation]{Notation}
\newtheorem{noterm}[equation]{Notation and Terminology}

\newcommand\define[1]{\emph{\textbf{#1}}}%italicize and bold-face %this seems like a good alternative

\numberwithin{equation}{section}

\newcommand{\be}{\begin{equation}}
\newcommand{\ee}{\end{equation}}
\newcommand{\bea}{\begin{eqnarray}}
\newcommand{\eea}{\end{eqnarray}}
\newcommand{\bx}{\begin{example}}
\newcommand{\ex}{\end{example}}
\newcommand{\bex}{\begin{exercise}}
\newcommand{\eex}{\end{exercise}}
\newcommand{\ban}{\begin{answer}}
\newcommand{\ean}{\end{answer}}
\newcommand{\bt}{\begin{theorem}}
\newcommand{\et}{\end{theorem}}
\newcommand{\bc}{\begin{corollary}}
\newcommand{\ec}{\end{corollary}}
\newcommand{\blem}{\begin{lemma}}
\newcommand{\elem}{\end{lemma}}
\newcommand{\bp}{\begin{problem}}
\newcommand{\ep}{\end{problem}}
\newcommand{\bn}{\begin{proposition}}
\newcommand{\en}{\end{proposition}}
\newcommand{\bd}{\begin{definition}}
\newcommand{\ed}{\end{definition}}
\newcommand{\bcon}{\begin{construction}}
\newcommand{\econ}{\end{construction}}
\newcommand{\bq}{\begin{question}}
\newcommand{\eq}{\end{question}}
\newcommand{\bprf}{\begin{proof}}
\newcommand{\eprf}{\end{proof}}
\newcommand{\br}{\begin{remark}}
\newcommand{\er}{\end{remark}}
\newcommand{\bs}{\begin{solution}}
\newcommand{\es}{\end{solution}}
\newcommand{\beqs}{\begin{eqnarray}}
\newcommand{\eeqs}{\end{eqnarray}}
\newcommand{\bnt}{\begin{noterm}}
\newcommand{\ent}{\end{noterm}}
\newcommand{\bnot}{\begin{notation}}
\newcommand{\enot}{\end{notation}}

\newcommand{\<}{\langle}
\renewcommand{\>}{\rangle}

\newcommand{\id}{\mathrm{id}}

\newcommand{\Tr}{{\rm Tr} }

\newcommand{\CPTP}{\mathbf{CPTP}}
\newcommand{\im}{\mathrm{im}}

\newcommand{\R}{\mathbb{R}}
\newcommand{\C}{\mathbb{C}}

\newcommand{\N}{\mathbb{N}}

\newcommand{\matr}{\mathbb{M}}%matrix algebra notation -- change as desired

\newcommand{\Jamiol}{\mathscr{J}}%for Jamiolkowski operator
\newcommand{\obs}[1]{\mathcal{O}_{#1}}%observable notation

\DeclareFontFamily{OT1}{pzc}{}
\DeclareFontShape{OT1}{pzc}{m}{it}{ <-> s*[1.2] pzcmi7t }{}
\DeclareMathAlphabet{\mathpzc}{OT1}{pzc}{m}{it}
\newcommand{\Alg}[1]{\mathpzc{#1}}
\newcommand{\map}[1]{\mathcal{#1}}

\newcommand{\ben}{\renewcommand{\theenumi}{\alph{enumi}} 
\renewcommand{\labelenumi}{(\theenumi)}\begin{enumerate}}
\newcommand{\een}{\end{enumerate}}

\newcommand\blfootnote[1]{%
  \begingroup
  \renewcommand\thefootnote{}\footnote{#1}%
  \addtocounter{footnote}{-1}%
  \endgroup
}

\title{\Large  Operator representation of spatiotemporal quantum correlations}
\author[1,2]{James Fullwood \thanks{fullwood@hainanu.edu.cn}}
\author[3]{Arthur J. Parzygnat \thanks{arthurjp@mit.edu}}
\affil[1]{School of Mathematics and Statistics, Hainan University, Haikou, Hainan, 570228, China}
\affil[2]{Hainan International Exchange Center for Theoretical Physics, Haikou, Hainan, 570228, China}
\affil[3]{Experimental Study Group, Massachusetts Institute of Technology, Cambridge, Massachusetts 02139, USA}
\date{}                     %% if you don't need date to appear
\date{October 10, 2026}

\begin{document}
\emergencystretch 2em

\maketitle

\begin{abstract}
While quantum correlations between two spacelike-separated systems are fully encoded by the bipartite density operator associated with the joint system, there does not exist an analogous operator representing general quantum correlations across space \emph{and} time. This is in stark contrast to the case of classical random variables, which make no distinction between spacelike and timelike correlations. Despite this, we show that spatiotemporal correlations between \emph{light-touch observables} (i.e., observables whose eigenvalues are all equal in magnitude) admit a \emph{unique} operator representation for arbitrary timelike-separated quantum systems. 
A special case of our result reproduces generalized Pauli observables and pseudo-density matrices, which have, up until now, only been defined for multi-qubit systems. In the case of qutrit systems, we use our results to illustrate an intriguing connection between light-touch observables and symmetric, informationally complete, positive operator-valued measures (SIC-POVMs). 

\end{abstract}

\vspace{-7mm}
\tableofcontents

%%%%%%%%%%%%%%%%%%%%%%%%
\section{Introduction}
%%%%%%%%%%%%%%%%%%%%%%%%

Quantum theory treats space and time in fundamentally different ways. In particular, correlations between spacelike-separated systems $A$ and $B$ are mathematically described by a unique density operator $\rho_{AB}$~\cites{vN18,NiCh11}, so that the expectation value $\<\obs{A}\otimes \obs{B}\>$ of the product of measurements of observables $\obs{A}$ and $\obs{B}$ is given by
\be \label{TFXSLSOXS71}
\<\obs{A}\otimes \obs{B}\>=\Tr\big[\rho_{AB}(\obs{A}\otimes \obs{B})\big]\, .
\ee
Moreover, the density operator $\rho_{AB}$ does not depend in any way on the measurements being made, so that equation \eqref{TFXSLSOXS71} holds \emph{for all} observables $\obs{A}$ and $\obs{B}$. 
On the other hand, if systems $A$ and $B$ correspond to the input and output of a quantum channel $\mathcal{E}$, which we view as governing the dynamics between a sequential measurement of $\obs{A}$ followed by $\obs{B}$, then the two-time expectation value $\langle \obs{A}\,,\obs{B} \rangle$ associated with the product of such a sequential measurement is given by
\be
\label{INDXOBX37}
\langle \mathcal{O}_{A}\, ,\mathcal{O}_{B} \rangle=\sum_{i}\lambda_i\Tr\big[\map{E}(P_i\rho P_i)\mathcal{O}_{B}\big]\, ,
\ee
where $\rho$ is the initial state of $A$ and $\obs{A}=\sum_{i}\lambda_i P_i$ is the spectral decomposition of $\obs{A}$~\cites{BCL90,Tammero18,FBC17,Fritz10}. 
However, there does not in general exist a spatiotemporal analogue $\varrho_{AB}$ of a density operator such that
\begin{equation} 
\label{eq:nogointro}
\langle \obs{A}\, ,\obs{B} \rangle=\Tr\big[\varrho_{AB}(\obs{A}\otimes\obs{B})\big]
\end{equation}
for all observables $\obs{A}$ and $\obs{B}$ (see Theorem~\ref{prop:nogolinearity}). As such, the operational interpretation of a bipartite density operator does not extend in a direct way to timelike-separated measurements%
%footnote
\footnote{In this paper, timelike separated will be taken to mean causally related by a channel. The two notions are, however, slightly different. See Ref.~\cite{LeSp13} for a discussion on the distinction between the two notions.},
%end footnote
thus resulting in a disparity in the mathematical tools one uses to investigate spatial versus temporal quantum correlations. 
In contrast, no such disparity arises in the setting of general relativity, where space, time, and causal structure are captured by a single entity \emph{spacetime}, which is mathematically modelled by a Lorentzian manifold~\cite{HaEl73}. 
So if we are to discover postulates for quantum gravity, it seems quite natural to seek a mathematical description of timelike-separated quantum systems that differs from the notion of a bipartite density operator in a manner that is more analogous to how spacetime differs from space~\cites{Ba06,Sorkin97,Rovelli01}. 

One approach towards such a framework is presented in Ref.~\cite{FJV15}, where the authors introduced a spatiotemporal quantum state encoding the two-time expectation values for all timelike-separated \emph{Pauli observables} on a system of qubits. The spatiotemporal state was termed a \emph{pseudo-density matrix}, where ``pseudo'' refers to that fact that, unlike density matrices, pseudo-density matrices are not positive in general, though they are still hermitian. 
In fact, the negative eigenvalues of a pseudo-density matrix serve as a witness to quantum correlations that imply causation~\cites{FJV15,ZPTGVF18,song_2023,LCD24,LJQLD24,JSK23}. 
Moreover, the associated pseudo-density matrix $\varrho_{AB}$ satisfies Equation~\eqref{eq:nogointro} for all \emph{Pauli observables} $\obs{A}$ and $\obs{B}$ on a system of \emph{qubits}. 
Therefore, pseudo-density matrices provide a spatiotemporal generalization of the operational interpretation of a bipartite density operator. However, the definition of a pseudo-density matrix relies heavily on several mathematical properties of Pauli observables. 
As such, the definition of a pseudo-density matrix had not been extended to systems of arbitrary dimension, and the question of whether or not a suitable generalization of pseudo-density matrix exists for arbitrary systems was an open question first posed in Ref.~\cite{HHPBS17}.

Following earlier results of Ref.~\cite{HHPBS17}, it was recently shown in Ref.~\cite{LQDV23} that if a system of qubits evolves according to a quantum channel $\map{E}$ between two timelike-separated measurements of Pauli observables, then the associated pseudo-density matrix $\varrho_{AB}$ may be given by the formula
\be \label{SMTXBLXS87}
\varrho_{AB}=\frac{1}{2}\big\{\rho\otimes \mathds{1}_{\!B}\, , \Jamiol[\map{E}]\big\}\, ,
\ee
where $\{\cdot\, , \cdot\}$ denotes the anti-commutator and $\Jamiol[\map{E}]=\sum_{i,j}|i\>\<j|\otimes \map{E}(|j\>\<i|)$ is the \emph{Jamio\l kowski matrix} of the channel $\map{E}$~\cite{Ja72}%
%footnote
\footnote{The Jamio\l kowski matrix $\Jamiol[\map{E}]$ is not to be confused with the Choi matrix of $\map{E}$, which is the partial transpose of $\Jamiol[\map{E}]$~\cite{Ch75}.}. 
%end footnote
As the right-hand-side (RHS) of Equation~\eqref{SMTXBLXS87} is defined for channels $\map{E}$ between arbitrary systems $A$ and $B$, the issue of generalizing pseudo-density matrices to arbitrary dimension may be reduced to the question of whether or not the operator corresponding to the RHS of Equation~\eqref{SMTXBLXS87} also admits an operational interpretation in terms of encoding two-time expectation values associated with measurements performed on arbitrary timelike-separated systems $A$ and $B$. 

In this work, we prove that the operator $\varrho_{AB}$ as given by~\eqref{SMTXBLXS87} indeed admits such an interpretation in terms of two-time expectation values $\<\obs{A}\,,\obs{B}\>$ for timelike-separated systems $A$ and $B$ of arbitrary dimension. 
More precisely, Theorem~\ref{MTXS45739} shows that for arbitrary timelike-separated systems $A$ and $B$, the operator $\varrho_{AB}$ as given by~\eqref{SMTXBLXS87} is the \emph{unique} operator satisfying Equation~\eqref{eq:nogointro} for all \emph{light-touch} observables $\obs{A}$ (i.e., observables with all eigenvalues equal in magnitude) and for all observables $\obs{B}$. Since Pauli operators and their tensor products are examples of light-touch observables, our theorem provides an extension of pseudo-density matrices to arbitrary finite-dimensional quantum systems (which retain their operational interpretation in terms of two-time expectation values), 
thus solving the open problem of finding such an extension~\cite{HHPBS17}. Moreover, when $\mathcal{E}$ is the identity channel, the operator $\varrho_{AB}$ coincides with the virtual broadcasting of the state $\rho_{A}$~\cite{PFBC23}. Hence, our results provide an operational meaning for the canonical virtual broadcasting map from Ref.~\cite{PFBC23} that is distinct from, but complementary to, the existing operational interpretation provided in Refs.~\cites{BDOV13,BDOV14} (see Remark~\ref{rmk:virtualbroadcasting} for further details). 

Another interesting aspect of the operator $\varrho_{AB}$ as given by~\eqref{SMTXBLXS87} is that it is an example of a \emph{quantum state over time}, a concept first defined in Ref.~\cite{HHPBS17} and later refined in Refs.~\cites{FuPa22,FuPa22a}. Similar to pseudo-density matrices, quantum states over time were first introduced as spatiotemporal generalizations of bipartite density operators that do not require any doubling of Hilbert spaces as in the process matrix formalism or related approaches~\cites{CDP09,OCB12,CJQW18,CRGWF18,RCGWF18,JiKa23}. It has since been shown that quantum states over time encompass many concepts introduced for similar purposes~\cite{FuPa22a}, such as quantum conditional operators~\cites{Le06,Le07}, quantum Bayesian inference~\cites{LeSp13,CoSp12,PaRuBayes,Ts22}, quantum causal models~\cite{ABHLS17}, multi-time states~\cites{Wat55,ABL64,ReAh95,APT10,APTV09,AhVa08}, past density matrices~\cites{GJM13,KhMo21,BJDJMSX20}, temporal density matrices~\cite{MiAdPr25}, two-point correlation functions~\cites{BDOV13,BDOV14,BHR17}, time-reversal symmetry~\cites{FuPa22a,PaBu22,PaFu24TSC}, and retrodiction~\cites{FuPa22a,PaBu22,Ts22,Ts22b,Pa24}. For example, multi-time states and weak values~\cites{ABL64,APTV09} are quantum states over time that have been used for the quantum tunneling time problem~\cites{Steinberg95,RSRS20}, quantum phase estimation~\cite{LYABPSH22}, aspects of pre- and post-selection~\cite{AHLLBL20}, a spacetime-symmetric formulation of phase space and Hamiltonian dynamics for quantization~\cite{DMR24}, time-like entanglement entropy~\cite{DHMTT22}, and the black hole information paradox~\cites{HoMa04,GoPr04}. 
Quantum states over time have also led to theoretical insights regarding dynamical measures of quantum information~\cites{FuPa23,MaCh22} and virtual quantum broadcasting~\cite{PFBC23}. 

While there have been various approaches to the construction of quantum states over time, Ref.~\cite{LiNg23} recently established that the assignment sending $(\map{E},\rho)$ to~\eqref{SMTXBLXS87}
defines the \emph{unique} quantum state over time satisfying a short list of physically motivated axioms. An alternative characterization of this assignment using axioms related to virtual quantum broadcasting was also recently proved by the authors, Buscemi, and Chiribella in Ref.~\cite{PFBC23}. 
As such, not only does the operator $\varrho_{AB}$ as given by \eqref{SMTXBLXS87} provide a unique operator representation of spatiotemporal correlations between light-touch observables, but in light of such characterization theorems, it may also be viewed as \emph{the canonical state over time} associated with the \emph{process} of $\rho$ evolving into $\map{E}(\rho)$ under the channel $\map{E}$. Moreover, as the axioms characterizing $\varrho_{AB}$ from the viewpoint of quantum states over time make no reference to two-time expectation values, it follows that the operator $\varrho_{AB}$ as given by \eqref{SMTXBLXS87} provides the unique answer to two separate fundamental questions related to spatiotemporal aspects of quantum information theory. It is for these reasons we expect the canonical state over time $\varrho_{AB}$ to play a fundamental role as we develop our understanding of quantum information and its relation to spacetime and quantum gravity. 

The paper is organized as follows. 
In Section~\ref{sec:TTEV}, we provide the precise definition of two-time expectation values $\<\obs{A},\obs{B}\>$ associated with a quantum system that evolves according to a quantum channel between measurements, and we prove various properties satisfied by such expectation values. 
In Section~\ref{sec:NGT}, we prove a no-go result (Theorem~\ref{prop:nogolinearity}) stating that for general timelike-separated systems $A$ and $B$, there does not exist an operator $\varrho_{AB}$ satisfying equation~\eqref{eq:nogointro} for \emph{all} observables $\obs{A}$ and $\obs{B}$. 
In Section~\ref{sec:PDMS}, we recall the definition of a pseudo-density matrix associated with timelike-separated systems of qubits, while setting the stage for its generalization to arbitrary systems. 
In Section~\ref{EXTOTX}, we define light-touch observables and analyze some of their properties before presenting our main result, Theorem~\ref{MTXS45739}, which provides an extension of pseudo-density matrices to quantum systems of arbitrary finite dimension, simultaneously providing an operational characterization theorem for the canonical quantum state over time from Ref.~\cite{FuPa22}. 
In Section~\ref{sec:qutrits}, we provide an illustration of our main results for qutrits (3-level systems), and we also show that in dimension 3 there exists a one-to-one correspondence between orthonormal bases of light-touch observables and symmetric, informationally complete, positive operator-valued measures (SIC-POVMs).

%%%%%%%%%%%%%%%%%%%%%%%%%%%%%%%%%%%%%%%%
\section{Two-time expectation values}
\label{sec:TTEV}
%%%%%%%%%%%%%%%%%%%%%%%%%%%%%%%%%%%%%%%%
In this section, we define a certain class of expectation values associated with two timelike-separated measurements on a quantum system and investigate certain properties they satisfy. The expectation values are associated with a system that first undergoes a projective measurement, then evolves according to a quantum channel, and then a second projective measurement is performed on the output of the channel. In later sections, we will show that such expectation values furnish an operational meaning for quantum states over time appropriately generalizing the operational meaning of a static quantum state. We first set some notation and terminology in place that will be used throughout this work.

\bnt
Throughout this work, $A$ and $B$ denote quantum systems with finite-dimensional Hilbert spaces $\mathcal{H}_A$ and $\mathcal{H}_B$. The caligraphic letters $\Alg{A}$ and $\Alg{B}$ will be used to denote the algebras of linear operators on $\mathcal{H}_A$ and $\mathcal{H}_B$, respectively, and the set of density matrices corresponding to states of $A$ will be denoted by $\mathcal{S}(A)$. We will often identify $\Alg{A}$ with the algebra $\mathbb{M}_d$ of $d\times d$ complex matrices whenever $d=\text{dim}(\mathcal{H}_A)$, and similarly for $\Alg{B}$. A linear map $\map{E}:\Alg{A}\to \Alg{B}$ which is completely positive and trace-preserving (CPTP) will be referred to as a \define{quantum channel}, and the set of all such quantum channels will be denoted by $\mathbf{CPTP}(A,B)$. Given a linear map $\map{E}:\Alg{A}\to\Alg{B}$, the \define{Jamio{\l}kowski matrix}~\cite{Ja72} of $\map{E}$ is the element $\mathscr{J}[\mathcal{E}]\in\Alg{A}\otimes\Alg{B}$ given by
\be \label{JAMIOXL71}
\mathscr{J}[\mathcal{E}]
=\sum_{i,j}|i\>\<j|\otimes \map{E}\big(|j\>\<i|\big)\, ,
\ee
where $\Alg{A}=\matr_{m}$, $\Alg{B}=\matr_{n}$, and $\{|i\>\}_{i=1}^{m}$ is the standard basis in $\mathcal{H}_{A}=\C^{m}$, so that $|i\>\<j|$ is the matrix with $1$ in the $ij$ entry and $0$ otherwise. 
The \define{Hilbert--Schmidt adjoint} $\map{E}^*:\Alg{B}\to\Alg{A}$ of a linear map $\map{E}:\Alg{A}\to\Alg{B}$ is the unique linear map satisfying
$
\Tr\big[\map{E}(X_{\!A})^{\dag}X_{\!B}\big]=\Tr\big[X_{\!A}^\dag\map{E}^*(X_{\!B})\big]
$
for all $X_{\!A}\in\Alg{A}$ and $X_{\!B}\in\Alg{B}$, where $\dag$ denotes the conjugate transpose. The set $\CPTP(A,B)\times \mathcal{S}(A)$ will be denoted by $\mathscr{P}(A,B)$, and an element $(\map{E},\rho)\in \mathscr{P}(A,B)$ will be referred to as a \define{process}. 
The set of distinct eigenvalues of an operator $X$ will be denoted by $\mathfrak{spec}(X)$, and Hermitian operators will be referred to as \define{observables}. The real vector space of all observables in $\Alg{A}$ will be denoted by $\mathbf{Obs}(A)$, and similarly for $\mathbf{Obs}(B)$. 
Given an observable $\obs{}$, the spectral decomposition 
\[
\obs{}=\sum_{\lambda\in \mathfrak{spec}(\obs{})}\lambda P_{\lambda}\, ,
\]
with $P_{\lambda}$ the orthogonal projection onto the $\lambda$-eigenspace of $\obs{}$, will be referred to as the \define{canonical spectral decomposition} of $\obs{}$. 
\ent

\bnot
The Pauli matrices are given by $\sigma_0=\mathds{1}_{2}$,
\[
\sigma_1=
\begin{pmatrix}0&1\\1&0\end{pmatrix},\,
 \sigma_2=\begin{pmatrix}0&-i \\i&0\end{pmatrix}, \text{ and }
 \sigma_3=\begin{pmatrix}1&0 \\0&-1 \\\end{pmatrix}.
\]
Given a 3-vector $\vec{x}=(x_1,x_2,x_3)\in\R^{3}$ and a $4$-vector $\mathbf{y}=(y_0,y_1,y_2,y_3)\in\R^{4}$, we set 
\[
\vec{x}\cdot\vec{\sigma}=\sum_{j=1}^{3}x_{j}\sigma_{j}=\left(
\begin{array}{cc}
x_3&x_1-ix_2 \\
x_1+ix_2&-x_3 \\
\end{array}
\right)
\]
and
\[
\mathbf{y}\cdot\boldsymbol{\sigma}=\sum_{j=0}^{3}y_{j}\sigma_{j}=\left(
\begin{array}{cc}
y_0+y_3&y_1-iy_2 \\
y_1+iy_2&y_0-y_3 \\
\end{array}
\right)\, .
\]
\enot

\bd
\label{defn:TTEV}
Let $\obs{A}\in \mathbf{Obs}(A)$ and $\obs{B}\in \mathbf{Obs}(B)$ be observables, with canonical spectral decompositions given by 
\be \label{SPXTRL357}
\obs{A}=\sum_{i=1}^{s}\lambda_iP_i \quad \text{and} \quad \obs{B}=\sum_{j=1}^{t}\mu_jQ_j\, ,
\ee
where $s$ and $t$ are the number of distinct eigenvalues of $\obs{A}$ and $\obs{B}$, respectively. 
Then the \define{two-time expectation value} of the observables $\obs{A}$ and $\obs{B}$ with respect to the process $(\map{E},\rho)\in \mathscr{P}(A,B)$ is the element $\langle \obs{A} \, , \obs{B}\rangle\in \R$ given by
\be \label{2TXEXT67}
\langle \obs{A} \, , \obs{B}\rangle=\sum_{i,j}\lambda_i\mu_j\Tr\big[\map{E}(P_i\rho P_i)Q_{j}\big]\, .
\ee
\ed

\br[The operational meaning of the two-time expectation values] \label{OPXINTXS347}
Given a process $(\map{E},\rho)\in \mathscr{P}(A,B)$ and projective measurements $\{P_i\}$ and $\{Q_j\}$ associated with the observables $\obs{A}$ and $\obs{B}$ as given by~\eqref{SPXTRL357}, the values 
\be \label{PLXSX37}
\mathbb{P}(i\,,j)=\Tr\big[\map{E}(P_i\rho P_i)Q_{j}\big]
\ee
define a probability distribution on $\{1,\ldots,s\}\times \{1,\ldots,t\}$. 
The probability distribution $\mathbb{P}(i\,,j)$ represents the probability of obtaining measurement outcomes $\lambda_i$ followed by $\mu_j$ when a system in an initial state $\rho\in \mathcal{S}(A)$ evolves according to the channel $\map{E}\in \CPTP(A,B)$ between the measurements~\cites{BCL90}. 
The particular form of the joint probability distribution $\mathbb{P}(i\,,j)$ as given by~\eqref{PLXSX37} utilizes the L{\"u}ders update rule, which yields a natural quantum instrument associated with the observable $\obs{A}$ due to its dependence only on the projectors onto the eigenspaces of $\obs{A}$~\cites{DaLe70,BLM1996,Lu06,BuLa2009}. 
The appearance of an instrument here is in contrast with the joint probabilities associated with measurements performed in parallel on spacelike-separated systems, as such probabilities are independent of the quantum instruments implementing the associated measurements. 
We also note that the probability distribution $\mathbb{P}(i\,,j)$ can be interpreted as the joint probability associated with the \emph{prior} probability 
\[
\mathbb{P}(i)=\Tr[P_{i}\rho P_{i}]=\Tr[\rho P_{i}]
\]
of measuring the outcome $\lambda_{i}$ and the \emph{conditional} probability 
\[
\mathbb{P}(j\,|\,i)=\Tr\left[\map{E}\left(\frac{P_{i}\rho P_{i}}{\mathbb{P}(i)}\right) Q_{j}\right]
\]
obtained by evolving the updated quantum state $P_{i}\rho P_{i}/\mathbb{P}(i)$ \cites{vN18,Lu06,FuPa22a} (provided that $\mathbb{P}(i)\ne0$) according to the channel $\map{E}$ and then measuring the outcome $\mu_{j}$. Indeed,
\[
\mathbb{P}(j\,|\,i)\mathbb{P}(i)=
\Tr\left[\map{E}\left(\frac{P_{i}\rho P_{i}}{\mathbb{P}(i)}\right) Q_{j}\right]\mathbb{P}(i)\\
=\Tr\big[\map{E}(P_i\rho P_i)Q_{j}\big]\\
=\mathbb{P}(i\,,j).
\]

Also, note that by summing over $j$ on the RHS of Equation~\eqref{2TXEXT67}, we arrive at the formula
\eqref{INDXOBX37}, 
which is more compact expression for $\langle \obs{A} \, , \obs{B}\rangle$ that will be used throughout. Finally, we note that the two-time expectation values have been given various names in the literature, including sequential measurement expectation values~\cite{BMKG13}, temporal averages~\cite{MVVAPGDG21}, quantum correlation functions~\cite{BTCV04},  correlators~\cite{Fritz10}, and two-point measurements~\cites{TLH07,LBLHFG23}.
\er

\br \label{STXTCVLX77}
Static (one-time) expectation values are special cases of two-time expectation values in the following sense. 
If $(\map{E},\rho)\in \mathscr{P}(A,B)$ is a process, then  
$\langle \obs{A} \, , \mathds{1}_{\!B}\rangle=\Tr[\rho\obs{A}]$ and $\<\mathds{1}_{\!A},\obs{B}\>
=\Tr\big[\map{E}(\rho)\obs{B}\big]$
for every observable $\obs{A}\in \mathbf{Obs}(A)$ and $\obs{B}\in\mathbf{Obs}(B)$.
\er

In the next example we show that, contrary to local measurements performed on spacelike-separated systems, the two-time expectation function $\<\cdot\,, \cdot\>$ is \emph{not} bilinear in general. 

\bx[Non-bilinearity of two-time expectation values]
\label{ex:nonlinearity}
Suppose $\Alg{A}=\Alg{B}=\matr_2$, let $(\map{E},\rho)\in \mathscr{P}(A,B)$ be the process given by
\[
(\map{E},\rho)=\left(\id_{\matr_{2}}\, ,|-\>\<-|\right)\, , \;\text{where}\; |-\>\<-|=\frac{1}{2}\begin{pmatrix}1&-1\\-1&1\end{pmatrix}\, .
\]
Now, given any 4-vector $\mathbf{x}=(x_0,\vec{x})\in\R^{4}$ with $\vec{x}=(x_1,x_2,x_3)$ satisfying $x_1^2+x_2^2>0$, then
\[
|\pm \hat{x}\>:=
\frac{1}{\sqrt{2\lVert\vec{x}\rVert(\lVert\vec{x}\rVert\pm x_{3})}}
\begin{pmatrix}x_{3}\pm\lVert\vec{x}\rVert\\x_{1}+ix_{2}\end{pmatrix}
\]
is a normalized eigenvector of $\vec{x}\cdot\vec{\sigma}$ with eigenvalue $\pm \lVert\vec{x}\rVert$. Its associated projection operator is
\[
|\pm \hat{x}\>\<\pm\hat{x}|=\frac{1}{2\lVert\vec{x}\rVert x_{3\pm}}\begin{pmatrix}x_{3\pm}^2&\pm \overline{x_{12}}x_{3\pm}\\\pm x_{12}x_{3\pm}&|x_{12}|^2\end{pmatrix},
\]
where 
\[
x_{12}:=x_{1}+ix_{2}
\;\text{ and }\;
x_{3\pm}:=\lVert\vec{x}\rVert\pm x_{3}.
\]
Furthermore,  $|\pm\hat{x}\>$ is an eigenvector of $\mathbf{x}\cdot\boldsymbol{\sigma}$ with eigenvalue $x_{0}\pm\lVert\vec{x}\rVert$. Thus, if we let $\mathbf{x}=(1,1,0,0)$, $\mathbf{y}=(-1,0,1,0)$, and $\mathbf{v}=\mathbf{x}-\mathbf{y}=(2,1,-1,0)$, then 
\[
|\pm\hat{x}\>=\frac{1}{\sqrt{2}}\begin{pmatrix}\pm 1\\1\end{pmatrix}=:\pm |\pm\>,
\quad
|\pm\hat{y}\>=\frac{1}{\sqrt{2}}\begin{pmatrix}\pm 1\\i\end{pmatrix}=:\pm|\pm i\>,
\quad\text{ and }\quad
|\pm\hat{v}\>=\frac{1}{2}\begin{pmatrix}\pm \sqrt{2}\\1-i\end{pmatrix}
\]
are the eigenvectors of $\obs{A}^{(1)}:=\mathbf{x}\cdot\boldsymbol{\sigma}$, $\obs{A}^{(2)}:=\mathbf{y}\cdot\boldsymbol{\sigma}$, and $\obs{A}^{(1)}-\obs{A}^{(2)}=\mathbf{v}\cdot\boldsymbol{\sigma}$, respectively, with respective eigenvalues  
\[
\lambda^{(1)}_{\pm}=1\pm 1\,,\quad
\lambda^{(2)}_{\pm}=-1\pm 1\,,\quad\text{ and }\quad
\lambda^{(12)}_{\pm}=2\pm \sqrt{2}\,.
\]
The projection operators associated with the canonical spectral decompositions of $\obs{A}^{(1)}$, $\obs{A}^{(2)}$ and $\obs{A}^{(1)}-\obs{A}^{(2)}$ are then
\[
P^{(1)}_{\pm}=\frac{1}{2}\begin{pmatrix}1&\pm 1\\\pm 1&1\end{pmatrix}\,,\quad
P^{(2)}_{\pm}=\frac{1}{2}\begin{pmatrix}1&\mp i\\\pm i&1\end{pmatrix}\,,
\quad\text{ and }\quad
P^{(12)}_{\pm}=\frac{1}{2\sqrt{2}}\begin{pmatrix}\sqrt{2}&\pm (1+ i)\\\pm(1-i)&\sqrt{2}\end{pmatrix}\,,
\]
respectively. Finally, taking $\obs{B}=(a,b,c,d)\cdot\boldsymbol{\sigma}$, one has 
\[
\Big\<\obs{A}^{(1)}- \obs{A}^{(2)},\obs{B}\Big\>
=a\,,
\]
while
\[
\Big\<\obs{A}^{(1)},\obs{B}\Big\>-\Big\<\obs{A}^{(2)},\obs{B}\Big\>
=0-(c-a)=a-c. 
\]
Thus, $\big\<\obs{A}^{(1)},\obs{B}\big\>-\big\<\obs{A}^{(2)},\obs{B}\big\>\neq \big\<\obs{A}^{(1)}- \obs{A}^{(2)},\obs{B}\big\>$ for all $c\neq 0$. As such, linearity of two-time expectation values in the first argument does not hold in general. 
\ex

Despite the lack of bilinearity of the two-time expectation value function for all processes, certain examples of processes $(\mathcal{E},\rho)$ admit linear representations.  
Interestingly, these examples correspond to when either there is complete ignorance of the initial state, i.e., when $\rho$ is maximally mixed, or when the system is reset after the first measurement, rendering subsequent measurement outcomes independent of the initial state, i.e., when $\mathcal{E}$ is a discard-and-prepare channel.

\bx \label{LMXS579}
Let $\Alg{A}=\matr_{m}$ and $\Alg{B}=\matr_{n}$ with $m,n\ge2$, and let $(\map{E},\rho)\in \mathscr{P}(A,B)$ be a process. Then the following statements hold.
\begin{enumerate}[i.]
\item \label{LMXS5791}
If $\rho=\frac{\mathds{1}_{m}}{m}$ is the maximally mixed state, then 
\[
\<\obs{A}\,, \obs{B}\>
=\Tr\left[\left(\frac{1}{m}\Jamiol[\map{E}]\right)(\obs{A}\otimes\obs{B})\right]
\]
for all observables $\obs{A}\in \mathbf{Obs}(A)$ and $\obs{B}\in\mathbf{Obs}(B)$, where $\mathscr{J}[\mathcal{E}]=\sum_{i,j}|i\>\<j|\otimes \mathcal{E}\big(|j\>\<i|\big)$. This follows from standard properties of the Jamio{\l}kowski matrix (see Refs.~\cites{Ja72,FuPa22a}). 

\item \label{LMXS5792}
If $\map{E}$ is a discard-and-prepare channel, i.e., there exists a state $\sigma\in\mathcal{S}(B)$ such that $\map{E}(X_{\!A})=\Tr[X_{\!A}]\,\sigma$ for all inputs $X_{\!A}\in\Alg{A}$, then 
\[
\<\obs{A}\,, \obs{B}\>
=\Tr\big[(\rho\otimes\sigma)(\obs{A}\otimes\obs{B})\big]
\]
for all observables $\obs{A}\in \mathbf{Obs}(A)$ and $\obs{B}\in\mathbf{Obs}(B)$.
\end{enumerate}
In either case, the two-time expectation value  $\langle\,\cdot\,, \,\cdot\, \rangle$ defines a real bilinear functional on $\mathbf{Obs}(A)\times\mathbf{Obs}(B)$. 
\ex

%%%%%%%%%%%%%%%%%%%%%%%%%%%%%%%%%%%%%%%%
\section{Non-representability of general processes}
\label{sec:NGT}
%%%%%%%%%%%%%%%%%%%%%%%%%%%%%%%%%%%%%%%%

In this section, we prove a no-go result stating that given arbitrary timelike-separated systems $A$ and $B$, there exists a process $(\mathcal{E},\rho)\in \mathscr{P}(A,B)$ such that the associated two-time expectation function $\<\,\cdot\,,\,\cdot\,\>$ cannot be represented by the trace of an operator $\varrho_{AB}$ acting on the tensor product of the associated observables. This is in stark contrast to the case when $A$ and $B$ are spacelike separated, since in such a case the correlation function between observables on $A$ and $B$ may always be represented by a bipartite density matrix $\rho_{AB}\in \Alg{A}\otimes \Alg{B}$. Thus, such a no-go theorem highlights a precise way in which quantum theory distinguishes between space and time.

\bd
A process $(\mathcal{E},\rho)\in \mathscr{P}(A,B)$ is said to be \define{representable} if and only if there exists an element $\varrho_{AB}\in \Alg{A}\otimes \Alg{B}$ such that for all observables $\mathcal{O}_{A}\in \mathbf{Obs}(A)$ and $\mathcal{O}_{B}\in \mathbf{Obs}(B)$,
$\<\mathcal{O}_{A}\, ,\mathcal{O}_{B}\>=\Tr\big[\varrho_{AB}(\mathcal{O}_{A}\otimes \mathcal{O}_{B})\big]$, 
where $\<\mathcal{O}_{A}\, ,\mathcal{O}_{B}\>$ is the two-time expectation value with respect to $(\mathcal{E},\rho)$. Otherwise, the process $(\mathcal{E},\rho)$ is said to be \define{non-representable}.
\ed

\bt[No-go Theorem for Operator Representations of Two-time Expectation Values] \label{prop:nogolinearity}
Let $\Alg{A}=\matr_{m}$ and $\Alg{B}=\matr_{n}$ with $m,n\ge 2$. 
Then there exists a process $(\mathcal{E},\rho)\in \mathscr{P}(A,B)$ that is non-representable.
\et

\bprf
It suffices to construct an example of a process $(\mathcal{E},\rho)\in \mathscr{P}(\mathbb{M}_m,\mathbb{M}_n)$ such that the two-time expectation function $\<\,\cdot\,,\,\cdot\,\>$ associated with $(\mathcal{E},\rho)$ is not bilinear. For the case $m=n=2$, such an example is provided by Example~\ref{ex:nonlinearity}. 

Now suppose $m,n$ are arbitrary, with $m,n\ge 2$. Although Example~\ref{ex:nonlinearity} was written in the special case of qubits, the example extends to arbitrary dimensions by appropriately restricting to the 2-dimensional subspace $\mathcal{H}\subseteq\C^{m}$ given by $\mathcal{H}=\mathrm{span}\big(e_{1}^{(m)},e_{2}^{(m)}\big)$, where $e^{(m)}_{1}$ and $e^{(m)}_{2}$ are the first two unit vectors in $\C^{m}$. Then set $\map{E}:\Alg{A}\to\Alg{B}$ to be the CPTP map given by 
\[
\map{E}(X_{\!A})=VX_{\!A}V^{\dag}+\Tr\big[P_{\mathcal{H}}^{\perp} X_{\!A}\big]\frac{\mathds{1}_{n}}{n}\,,
\quad\text{ where} \quad
V=\begin{pmatrix}\mathds{1}_{2}&0\\0&0\end{pmatrix}
\]
is an $n\times m$ partial isometry that sends the subspace $\mathcal{H}$ to the 2-dimensional subspace $\mathcal{K}=\mathrm{span}\big(e^{(n)}_{1},e^{(n)}_{2}\big)\subseteq\C^{n}$ isometrically and sends the orthogonal complement $\mathcal{H}^{\perp}\subseteq\C^{m}$ to the zero subspace in $\C^{n}$. Here, $P_{\mathcal{H}}=V^{\dag}V\in\Alg{A}$ is the orthogonal projection operator onto the subspace $\mathcal{H}$ inside $\C^{m}$, while $P_{\mathcal{H}}^{\perp}=\mathds{1}_{m}-P_{\mathcal{H}}$ is the orthogonal projection onto $\mathcal{H}^{\perp}$. Example~\ref{ex:nonlinearity} can now be applied on $\mathcal{H}$ and $\mathcal{K}$, since $\mathcal{H}\oplus\mathcal{H}^{\perp}=\C^{m}$ and $\mathcal{K}\oplus\mathcal{K}^{\perp}=\C^{n}$. 
More specifically, take
\[
\obs{\Alg{A}}^{(1)}=\begin{pmatrix}\mathbf{x}\cdot\boldsymbol{\sigma}&0\\0&(x_0+\lVert\vec{x}\rVert)\mathds{1}_{m-2}\end{pmatrix},
\quad
\obs{\Alg{A}}^{(2)}=\begin{pmatrix}\mathbf{y}\cdot\boldsymbol{\sigma}&0\\0&(y_0+\lVert\vec{y}\rVert)\mathds{1}_{m-2}\end{pmatrix},
\quad
\rho=\begin{pmatrix}|-\>\<-|&0\\0&0\end{pmatrix},
\]
and $\obs{B}=(a,b,c,d)\cdot\boldsymbol{\sigma}$, 
where $\mathbf{x}$, $\mathbf{y}$, and $|-\>\<-|$ are defined as in Example~\ref{ex:nonlinearity} (the choices of the $(m-2)\times(m-2)$ bottom-right subblocks of $\obs{A}^{(1)}$ and $\obs{A}^{(2)}$ guarantee that the spectra are the same as in Example~\ref{ex:nonlinearity}). 
\eprf

The fact that there exists non-representable processes implies that there are two-time quantum correlations which may not be realized by spacelike-separated systems, whose correlations are always representable by a density operator. 
Despite having only constructed one example of a non-representable process in all dimensions, we suspect that generic processes $(\mathcal{E},\rho)$ are non-representable. 
The only exceptions we have found so far occur when either $\mathcal{E}$ is a discard-and-prepare channel or $\rho$ is the maximally mixed state (cf.\ Example~\ref{LMXS579}). We leave this as an open question for future study. 

%%%%%%%%%%%%%%%%%%%
\section{Pseudo-density matrices}
\label{sec:PDMS}
%%%%%%%%%%%%%%%%%%%
In light of the no-go result of the previous section, one should not expect to obtain an operator representation in $\Alg{A}\otimes\Alg{B}$ of the two-time expectation function $\<\,\cdot\,,\,\cdot\,\>$ with respect to general processes $(\mathcal{E},\rho)\in\mathscr{P}(A,B)$. However, the pseudo-density matrix construction yields an operator representation for two-time expectation functions restricted to the set of Pauli observables for systems of qubits, which we now recall. 

Suppose $A$ and $B$ each consist of $m$ qubits, so that $\Alg{A}=\Alg{B}=\matr_{2^{m}}=\matr_{2}^{\otimes m}$. For every $\alpha\in \{0,\ldots,3\}^{m}$, we let $\alpha_{j}\in \{0,\ldots,3\}$ denote the $j^{\text{th}}$ component of $\alpha$ for all $j\in \{1,\ldots,m\}$, and we let $\sigma_{\alpha}\in \matr_{2}^{\otimes m}$ be the element given by
$
\sigma_{\alpha}=\sigma_{\alpha_1}\otimes \cdots \otimes \sigma_{\alpha_m}\, ,
$
where $\sigma_{i}$ denotes the $i^{\text{th}}$ Pauli matrix for $i\in \{0,\ldots,3\}$. If $A$ and $B$ are \emph{spacelike}-separated, then any bipartite density matrix $\rho_{AB}\in \Alg{A}\otimes\Alg{B}$ admits the expansion
\be \label{EXPXRHX37}
\rho_{AB}=\frac{1}{4^m}\sum_{\alpha,\beta\in \{0,\ldots,3\}^{m}}\<\sigma_{\alpha}\otimes \sigma_{\beta}\>\,\sigma_{\alpha}\otimes \sigma_{\beta}\, ,
\ee
where $\<\sigma_{\alpha}\otimes \sigma_{\beta}\>=\Tr\big[\rho_{AB} (\sigma_{\alpha}\otimes \sigma_{\beta})\big]$ is the expectation value of the product Pauli observable $\sigma_{\alpha}\otimes \sigma_{\beta}$ with respect to $\rho_{AB}$. However, if $A$ and $B$ represent the \emph{same} system at two times $t_A$ and $t_B>t_A$, then extending the RHS of \eqref{EXPXRHX37} to the setting of two \emph{timelike}-separated Pauli measurements on a single system of $m$ qubits motivates the following definition~\cite{FJV15}. 

\bd \label{PDMDFXSRSX67}
The \define{pseudo-density matrix} associated with a process $(\map{E},\rho)\in \mathscr{P}(A,B)$, where $\Alg{A}=\Alg{B}=\matr_{2^{m}}$, is the element $R(\map{E},\rho)\in \Alg{A}\otimes\Alg{B}$ given by
\[
R(\map{E},\rho)=\frac{1}{4^m}\sum_{\alpha,\beta\in \{0,\ldots,3\}^{m}}\<\sigma_{\alpha}\, ,\sigma_{\beta}\>\,\sigma_{\alpha}\otimes \sigma_{\beta}\, ,
\]
where $\<\sigma_{\alpha}\, ,\sigma_{\beta}\>$ is the two-time expectation value with respect to $(\map{E},\rho)$.
\ed

\bn \label{UNXPXM687}
Let $(\map{E},\rho)\in \mathscr{P}(A,B)$ be an $m$-qubit process, so that $\Alg{A}=\Alg{B}=\matr_{2^m}$. Then the pseudo-density matrix $R(\map{E},\rho)\in \Alg{A}\otimes\Alg{B}$ is the unique matrix such that
\be \label{EXTPDMXS737}
\<\sigma_{\gamma}\, , \sigma_{\delta}\>=\Tr\big[R(\map{E},\rho)(\sigma_{\gamma}\otimes \sigma_{\delta})\big]\, 
\ee
for all $\gamma,\delta\in \{0,\ldots , 3\}^{m}$. Moreover, $R(\map{E},\rho)$ is hermitian. 
\en

\bprf
The statement is a consequence of the fact that the set of all tensor products $\sigma_{\gamma}\otimes\sigma_{\delta}$ of Pauli matrices is an orthogonal basis for $\Alg{A}\otimes\Alg{B}$, where the orthogonality is defined with respect to the Hilbert--Schmidt inner product. See Ref.~\cite{FJV15} and the proof of Theorem~\ref{MTXS45739} for more details. 
\eprf

Although pseudo-density matrices have been defined for multi-qubit systems, whether or not pseudo-density matrices can be extended beyond qubit systems was an open question first posed in Ref.~\cite{HHPBS17}. In particular, what replaces Pauli observables in arbitrary dimensions for defining $R(\mathcal{E},\rho)$? 
Pauli observables satisfy many convenient properties; they are hermitian, unitary, traceless (except for the identity), they form an orthogonal real basis for observables with respect to the Hilbert--Schmidt inner product, and they have eigenvalues either all $+1$ or $\pm 1$. It is not possible to find such bases satisfying all of these properties in arbitrary dimensions. 

On the other hand, Refs.~\cites{HHPBS17,LQDV23} showed that the pseudo-density matrix $R(\map{E},\rho)$
may be given by the expression $\frac{1}{2}\big\{\rho\otimes\mathds{1}_{\!B},\Jamiol[\mathcal{E}]\big\}$ in the RHS of~\eqref{SMTXBLXS87}. 
As this expression is well-defined for arbitrary finite-dimensional quantum systems, one may use it to define a generalization of pseudo-density matrices outside the context of qubits~\cite{FuPa22}, provided that it retains a suitable operational and physically meaningful interpretation. 
In fact, the expression defines a \emph{quantum state over time} associated with the process $(\map{E},\rho)$, a notion that was first formulated in Ref.~\cite{HHPBS17} as a spatiotemporal generalization of a density matrix and later refined in Refs.~\cites{FuPa22,FuPa22a}. 
Moreover, the particular quantum state over time given by $\frac{1}{2}\big\{\rho\otimes\mathds{1}_{\!B},\Jamiol[\mathcal{E}]\big\}$ was characterized in Refs.~\cites{LiNg23,PFBC23} by a list of physically meaningful axioms, thus motivating the following definition. 

\bd
\label{defn:CSOT}
The \define{canonical state over time} associated with the process  $(\map{E},\rho)\in \mathscr{P}(A,B)$ is the element $\mathcal{E}\star \rho\in \Alg{A}\otimes\Alg{B}$ given by $\map{E}\star\rho=\frac{1}{2}\big\{\rho\otimes\mathds{1}_{\!B},\Jamiol[\map{E}]\big\}$. 
\ed

In the next section, we show that the canonical state over time is the unique operator representation of two-time expectation functions restricted to 
\emph{light-touch observables}, which provide a suitable generalization of Pauli observables in all dimensions. Such a result will then provide an affirmative answer to the open question of whether or not there exists a physically meaningful extension of the notion of pseudo-density matrix beyond systems of qubits.

%%%%%%%%%%%%%%%%%%%%%%%%%%%%%%%%%%%%%%%%%%%
\section{Operator representation for light-touch observables}
\label{EXTOTX}
%%%%%%%%%%%%%%%%%%%%%%%%%%%%%%%%%%%%%%%%%%%

In this section, we state and prove Theorem~\ref{MTXS45739}, which generalizes the characterization of pseudo-density matrices as the unique matrices encoding two-time expectation values associated with Pauli observables on a system of qubits (Proposition~\ref{UNXPXM687}). Not only does our theorem provide an extension of the uniqueness result for pseudo-density matrices to \emph{all} finite-dimensional quantum systems, but it also provides an operational meaning and characterization for canonical quantum states over time in terms of the expectation values of sequential measurements of light-touch observables. 
This solves the open problem of extending pseudo-density matrices to arbitrary quantum systems, it isolates the essential features of Pauli observables needed for such an extension, and it shows how quantum states over time provide a canonical framework for analyzing such spatiotemporal correlations in quantum physics.

Motivated by the terminology used in Ref.~\cite{LQDV23}, we introduce the following definition. 

\bd
\label{defn:lighttouch}
An observable $\obs{}$ will be referred to as \define{light touch} if and only if all of its eigenvalues have the same magnitude. 
\ed

\bx
The set of Pauli observables for multi-qubit systems are all light-touch observables. 
Light-touch observables share many common properties with Pauli observables. 
For example, any light-touch observable that is not the zero matrix is automatically unitary up to a scalar. Namely, if $\obs{}$ is a light-touch observable with spectrum $\{\pm \lambda\}$, where $\lambda>0$, then $\frac{1}{\sqrt{\lambda}}\obs{}$ is unitary. 
Examples of light-touch observables in dimension 3 will be given in Section~\ref{sec:qutrits}. 
\ex

The set of light-touch observables is informationally complete in the following sense. 

\bn
\label{prop:LTIC}
For any finite-dimensional quantum system $A$, there exists a basis of light-touch observables for $\mathbf{Obs}(A)$.
\en

\bprf 
The main step will be to show that the real-linear span of light-touch observables is all of $\mathbf{Obs}(A)$. 
The rest follows from the existence of a basis of $\mathbf{Obs}(A)$ within this set~\cite{Halmos1958}. 
To see the spanning condition, let $P$ be an arbitrary projection in $\Alg{A}$. Then  
$P=\frac{1}{2}(P-P^{\perp})+\frac{1}{2}\mathds{1}_{\!A}$
shows that every projection is a linear combination of one or two light-touch observables. Hence, since every observable is a linear combination of projections by the spectral theorem, this proves that light-touch observables span $\mathbf{Obs}(A)$, thus completing the proof.
\eprf

The next theorem is the main result of this work, which is a representation theorem for two-time expectation functions restricted to light-touch observables between arbitrary causally-related, timelike-separated quantum systems.

\bt[Operator Representation for Light-touch Two-time Expectation Values] 
\label{MTXS45739}
Let $(\map{E},\rho)\in \mathscr{P}(A,B)$ be a process.
Then there exists a unique element 
$\varrho_{AB}\in\Alg{A}\otimes\Alg{B}$ such that 
\be \label{EXTVISQSOT3791}
\<\obs{A}\, ,\obs{B}\>=\Tr\big[\varrho_{AB}(\obs{A}\otimes \obs{B})\big]
\ee
for all light-touch observables $\obs{A}\in \mathbf{Obs}(A)$ and for all observables $\obs{B}\in\mathbf{Obs}(B)$. Moreover, $\varrho_{AB}=\mathcal{E}\star \rho$, i.e., $\varrho_{AB}$ is the canonical state over time (as in Definition~\ref{defn:CSOT}). 
\et

\bprf
We first note that 
\be \label{TFX87}
\Tr_{A}\big[\Jamiol[\map{E}](X_{\!A}\otimes X_{\!B})\big]=\map{E}(X_{\!A})X_{\!B}\, 
\ee
for all $X_{\!A}\in\Alg{A}$ and $X_{\!B}\in\Alg{B}$, which is standard~\cites{Ja72,FuPa22a}. 

We now prove that $\varrho_{AB}=\map{E}\star\rho$ from Definition~\ref{defn:CSOT} satisfies~\eqref{EXTVISQSOT3791} (we will prove uniqueness afterwards). There are two cases to consider based on whether the observable $\obs{A}$ has a single eigenvalue or two. 

If $\mathfrak{spec}(\obs{A})=\{\lambda\}$, then $\obs{A}=\lambda\mathds{1}$. Therefore, 
\begingroup
\allowdisplaybreaks
\begin{align*}
\<\obs{A},\obs{B}\>&=\lambda\Tr\big[\map{E}(\rho)\obs{B}\big]\\
&=\lambda\Tr\Big[\Tr_{A}\big[\Jamiol[\map{E}](\rho\otimes\obs{B})\big]\Big]\\
&=\frac{\lambda}{2}\Tr\Big[\big\{\rho\otimes\mathds{1}_{\!B},\Jamiol[\map{E}]\big\}(\mathds{1}_{\!A}\otimes\obs{B})\Big]\\
&=\Tr\big[\varrho_{AB}(\obs{A}\otimes \obs{B})\big].
\end{align*}
\endgroup
The first equality holds by Remark~\ref{STXTCVLX77}, the second by~\eqref{TFX87}, the third by the properties of the trace, and the fourth by the definition of $\varrho_{AB}$. 

Now suppose $\mathfrak{spec}(\obs{A})=\{\pm \lambda\}$ for some $\lambda>0$. It then follows that $\obs{A}=\lambda(P_{1}-P_{2})$, where $P_1$ and $P_2$ are the orthogonal projections onto the $\pm\lambda$ eigenspaces. Since $\mathds{1}_{\!A}=P_{1}+P_{2}$, we have
\[
\lambda\mathds{1}_{\!A}+\obs{A}=2 \lambda P_{1} 
\quad\text{ and }\quad 
\lambda \mathds{1}_{\!A}-\obs{A}=2\lambda P_{2}\, ,
\]
which yields 
\[
\lambda P_1\rho P_1 - \lambda P_2\rho P_2
=\frac{1}{4\lambda}(\lambda\mathds{1}_{\!A}+\obs{A})\rho(\lambda\mathds{1}_{\!A}+\obs{A})-\frac{1}{4\lambda}(\lambda\mathds{1}_{\!A}-\obs{A})\rho(\lambda\mathds{1}_{\!A}-\obs{A}) =\frac{1}{2}\big\{\rho\, ,\obs{A}\big\}\, .
\]
We then have
\begingroup
\allowdisplaybreaks
\begin{eqnarray*}
\langle \obs{A} \, , \obs{B}\rangle&\!\!\!=\!\!\!&\sum_{i}\lambda_i\Tr\big[\map{E}(P_{i}\rho P_{i})\obs{B}\big] \\
&\!\!\!\overset{\eqref{TFX87}}=\!\!\!&\sum_{i}\lambda_i \Tr\Big[\Jamiol[\map{E}]\big((P_{i}\rho P_{i})\otimes \obs{B}\big)\Big] \\
&\!\!\!=\!\!\!&\Tr\left[\Jamiol[\map{E}]\Big(\big(\lambda P_1\rho P_1-\lambda P_2\rho P_2\big)\otimes \obs{B}\Big)\right] \\
&\!\!\!=\!\!\!&\Tr\left[\Jamiol[\map{E}]\left(\left(\frac{1}{2}\big\{\rho\, , \obs{A}\big\}\right)\otimes \obs{B}\right)\right] \\
&\!\!\!=\!\!\!&\frac{1}{2}\Tr\Big[\big\{\rho\otimes \mathds{1}_{\!B},\Jamiol[\map{E}]\big\}(\obs{A}\otimes \obs{B})\Big] \\
&\!\!\!=\!\!\!&\Tr\big[\varrho_{AB}(\obs{A}\otimes \obs{B})\big]\,.
\end{eqnarray*}
\endgroup
Thus, $\varrho_{AB}=\map{E}\star\rho$ satisfies~\eqref{EXTVISQSOT3791}. 

Uniqueness of $\varrho_{AB}$ follows from Proposition~\ref{prop:LTIC}, namely that there exists a basis of light-touch observables for $\mathbf{Obs}(A)$. In more detail, if $\{W_{\alpha}\}$ is such a basis for $\mathbf{Obs}(A)$ and if $\{Z_{\beta}\}$ is any basis of $\mathbf{Obs}(B)$, then the collection $\{W_{\alpha}\otimes Z_{\beta}\}$ forms a basis of $\Alg{A}\otimes\Alg{B}$. Therefore, if $X\in\Alg{A}\otimes\Alg{B}$ is some element satisfying $\<\obs{A}\, ,\obs{B}\>={\Tr\big[X(\obs{A}\otimes\obs{B})\big]}$
for all light-touch observables $\obs{A}\in\Alg{A}$ and for all observables $\obs{B}\in\Alg{B}$, then setting $Y=\map{E}\star\rho-X$ yields
$\Tr\big[Y(W_{\alpha}\otimes Z_{\beta})\big]=0$
for all $\alpha$ and $\beta$. Since the trace is faithful (i.e., the Hilbert--Schmidt inner product is non-degenerate), this implies $Y=0$, i.e., $X=\map{E}\star\rho$. 
\eprf

Theorem~\ref{MTXS45739} provides further justification for the terminology ``canonical state over time'' for $\map{E}\star \rho$, as it provides an operational characterization of this state over time, supplementing alternative characterizations~\cites{PFBC23,LiNg23}.
Moreover, since Pauli observables are light-touch observables, it immediately follows from Theorem~\ref{MTXS45739} and Proposition~\ref{UNXPXM687} that pseudo-density matrices are in fact canonical states over time, a result that was first proved in Ref.~\cite{HHPBS17} for qubits and later in Ref.~\cite{LQDV23} for multi-qubit systems. More generally, if $\{L_{\alpha}^{A}\}$ and $\{L_{\beta}^{B}\}$ are \emph{orthonormal} bases of light-touch observables for $\mathbf{Obs}(A)$ and $\mathbf{Obs}(B)$, respectively, it follows that ${\mathcal{E}\star\rho}=\sum_{\alpha,\beta}\<L_{\alpha}^{A},L_{\beta}^{B}\>L_{\alpha}^{A}\otimes L_{\beta}^{B}$, thereby extending the formula for the pseudodensity matrix from Definition~\ref{PDMDFXSRSX67} to arbitrary quantum systems. 
Since the Wootters phase-point operators provide such a basis for any prime dimension~\cite{Wootters_1987}, one obtains an orthogonal basis of light-touch observables in arbitrary finite dimension by taking tensor products of Wootters phase-point operators (see also Section~\ref{sec:qutrits} for an alternative construction for qutrit systems).

It also turns out that the set of light-touch observables is maximal in the sense that it is precisely the set of observables whose temporal correlations are represented by the canonical state over time $\map{E}\star\rho$ for all processes $(\map{E},\rho)\in\mathscr{P}(A,B)$.

\bt
\label{thm:maximal}
%Let $A$ be a quantum system represented by $\Alg{A}=\matr_{m}$.
Let $\Alg{M}\subseteq\mathbf{Obs}(A)$  be the largest subset satisfying
$\<\obs{A}\, ,\obs{B}\>=\Tr\big[(\map{E}\star\rho)(\obs{A}\otimes \obs{B})\big]$ for all $\obs{A}\in\Alg{M}$, all $\obs{B}\in\mathbf{Obs}(B)$, all processes $(\map{E},\rho)\in \mathscr{P}(A,B)$, and all systems $B$, where $\map{E}\star\rho$ is as in Definition~\ref{defn:CSOT}. Then $\Alg{M}$ is the set of all light-touch observables in $\mathbf{Obs}(A)$. 
\et

\bprf
It follows from Theorem~\ref{MTXS45739} that $\Alg{M}$ contains the set of light-touch observables in $\mathbf{Obs}(A)$. Thus, it remains to show that if $\obs{A}\in\Alg{M}$, then $\obs{A}$ is a light-touch observable. Write $\obs{A}=\sum_{k}\lambda_{k}P_{k}$ as its canonical spectral decomposition. 
Then
\begingroup
\allowdisplaybreaks
\begin{eqnarray*}
\sum_{k}\lambda_{k}\Tr\big[P_{k}\rho P_{k}\map{E}^{*}(\obs{B})\big]
&\!\!\!=\!\!\!&\sum_{k}\lambda_{k}\Tr\big[\map{E}(P_{k}\rho P_{k})\obs{B}\big] \\
&\!\!\!=\!\!\!&\<\obs{A},\obs{B}\>\\
&\!\!\!=\!\!\!&\sum_{k}\lambda_{k}\Tr\big[(\map{E}\star\rho)(P_{k}\otimes\obs{B})\big] \\
&\!\!\!=\!\!\!&\sum_{k}\lambda_{k}\Tr\left[\frac{1}{2}\big\{P_{k}\, ,\rho\big\}\,\map{E}^{*}(\obs{B})\right]
\end{eqnarray*}
\endgroup
for all observables $\obs{B}\in\mathbf{Obs}(B)$, density matrices $\rho\in\mathcal{S}(A)$, CPTP maps $\map{E}:\Alg{A}\to\Alg{B}$, and all $B$. The first equality follows by the definition of $\map{E}^*$, the second by the definition of $\<\,\cdot\,,\,\cdot\,\>$, the third by the definition of $\Alg{M}$, and the fourth from calculations similar to those in the proof of Theorem~\ref{MTXS45739}. Therefore, 
\[
\Tr\left[\sum_{k}\lambda_{k}\left(P_{k}\rho P_{k}-\frac{1}{2}\big\{P_{k}\,,\rho\big\}\right)\map{E}^{*}(\obs{B})\right]=0
\]
for all observables $\obs{B}\in\mathbf{Obs}(B)$, CPTP maps $\map{E}:\Alg{A}\to\Alg{B}$, and all $B$. By the faithfulness of the trace, this implies 
\[
\sum_{k}\lambda_{k}\left(P_{k}\rho P_{k}-\frac{1}{2}\big\{P_{k}\,,\rho\big\}\right)=0
\]
for all density matrices $\rho\in\mathcal{S}(A)$. This last expression is equivalent to
\be
\label{eq:projectoffdiagterms}
\sum_{k}\lambda_{k}\left(P_{k}\rho P_{k}^{\perp}+P_{k}^{\perp}\rho P_{k}\right)=0
\ee
for all density matrices $\rho\in\mathcal{S}(A)$. We now analyze the allowed $\obs{A}$ satisfying~\eqref{eq:projectoffdiagterms} depending on the number of terms in its spectral decomposition. 
For concreteness, we set $\Alg{A}=\matr_{m}$.

First, suppose $\obs{A}$ has none or only a single projector in its spectral decomposition, so that $\obs{A}=0$ or $\obs{A}=\lambda P$ for some projector $P$. If $\obs{A}=0$ or $P=\mathds{1}_{m}$, then $\obs{A}$ is a light-touch observable, which proves the claim. If $P$ is such that $P$ and $P^{\perp}$ are both nontrivial (neither $0$ nor $\mathds{1}_{m}$), then \eqref{eq:projectoffdiagterms} cannot hold for all density matrices $\rho\in\mathcal{S}(A)$, which leads to a contradiction. To see this, take any unit vectors $|\phi\>\in \im(P)$ and $|\psi\>\in \im(P^{\perp})$, where $\im(\,\ast\,)$ denotes the image, define $|\eta\>=\big(|\phi\>+|\psi\>\big)/\sqrt{2}$, and set $\rho=|\eta\>\<\eta|$. Then 
\[
\sum_{k}\lambda_{k}\left(P_{k}\rho P_{k}^{\perp}+P_{k}^{\perp}\rho P_{k}\right)
=
\lambda\big(P\rho P^{\perp}+P^{\perp}\rho P\big)=\frac{\lambda}{2}\big(|\phi\>\<\psi|+|\psi\>\<\phi|\big),
\]
which is not zero since $\lambda\ne0$. This shows that if $\obs{A}$ has only one nontrivial projector (neither $0$ nor $\mathds{1}_{m}$) in its spectral decomposition, then $\obs{A}$ cannot be an element of $\Alg{M}$.

Second, if $\obs{A}$ has a spectral decomposition with at least two projectors, i.e., at least two non-vanishing eigenvalues, then consider two distinct indices $i,j$, with corresponding distinct eigenvalues $\lambda_{i},\lambda_{j}$. Multiplying \eqref{eq:projectoffdiagterms} on the left by $P_{i}$ and the right by $P_{j}$ then gives
\[
0=P_{i}\sum_{k}\lambda_{k}\left(P_{k}\rho P_{k}^{\perp}+P_{k}^{\perp}\rho P_{k}\right)P_{j}
=\sum_{k}\lambda_{k}\delta_{ik}P_{k}\rho P_{k}^{\perp}P_{j}+\sum_{k}\lambda_{k}\delta_{kj}P_{i}P_{k}^{\perp}\rho P_{k}
=(\lambda_{i}+\lambda_{j})P_{i}\rho P_{j}
\]
for all density matrices $\rho\in\mathcal{S}(A)$. The only way this equality holds for all density matrices $\rho$ is if $\lambda_{j}=-\lambda_{i}$. Since $i$ and $j$ were arbitrary distinct indices in the spectral decomposition of $\obs{A}$, this requires that exactly two such indices can exist, and therefore, $\obs{A}$ must be a light-touch observable. This proves that $\Alg{M}$ is contained in the set of light-touch observables, thus, $\Alg{M}$ is equal to the set of light-touch observables in $\Alg{A}$.
\eprf

\br[Operational interpretation of virtual broadcasting]
\label{rmk:virtualbroadcasting}
A direct consequence of Theorem~\ref{MTXS45739} is that it provides an operational interpretation of the notion of \emph{virtual quantum broadcasting} recently put forward in Ref.~\cite{PFBC23}. By definition, the virtual broadcasting map of Ref.~\cite{PFBC23} is characterized by a simple list of axioms for broadcasting but dropping the requirement of complete positivity. As a function, it sends an input state $\rho$ to the canonical state over time $\id\star\rho$, which is not positive in general. The lack of positivity prevents the virtual broadcasting map from being a quantum channel, and as such, may not be considered as a physical quantum operation in the traditional sense. However, Refs.~\cites{BDOV13,BDOV14} showed how the virtual broadcasting map retains an operational meaning by providing an optimal statistical decomposition of it in terms of symmetric and anti-symmetric optimal cloners~\cite[Proposition~3]{BDOV13}. Moreover, Theorem~\ref{MTXS45739} yields a new operational meaning for virtually broadcasted states in terms of encoding two-time expectation values of light-touch observables, thus providing an alternative concrete physical interpretation for the virtual broadcasting map.
\er

%%%%%%%%%%%%%%%%%
\section{Example for qutrits}
\label{sec:qutrits}
%%%%%%%%%%%%%%%%%

In this section, we work out an explicit example for qutrits, i.e., $\Alg{A}=\Alg{B}=\matr_{3}$. In particular, we first construct an explicit informationally complete set of light-touch observables, i.e., a basis of $\mathbf{Obs}(A)$ consisting of light-touch observables. We then use this basis to construct a pseudo-density matrix for qutrits and illustrate how this agrees with the canonical state over time from Definition~\ref{defn:CSOT}. It turns out that for qutrits, there is a correspondence between an orthonormal set of light-touch observables and symmetric informationally complete positive operator-valued measures (SIC-POVMs), whose definition we now recall~\cites{Caves1999SICPOVMs,CFS02,FuSa03,RBKSC04}. 

\bd
Fix a dimension $d\in\N$. A \define{symmetric informationally complete positive operator-valued measure} (SIC-POVM) consists of a set of $d^2$ rank-$1$ orthogonal projections $P_{j}$ satisfying 
$\Tr[P_{j}P_{k}]=\frac{1}{d+1}$ 
for all $j\ne k$. The associated POVM is given by the set $\{P_{j}/d\}$. 
\ed

It is a consequence of the definition that $\frac{1}{d}\sum_{j=1}^{d^2}P_{j}=\mathds{1}_{d}$ and that the set $\{P_{j}\}$ forms a basis of all observables in $\matr_{d}$ (see Ref.~\cite{RBKSC04} for details). 

\bn
\label{prop:LTOSICPOVM}
Let $\{P_{j}\}$ be any SIC-POVM in $\matr_{d}$ with $d=3$. Then the operators 
$L_{j}:=2P_{j}-\mathds{1}_{d}\equiv P_{j}-P_{j}^{\perp}$
are all light-touch observables and the set $\big\{L_{j}/\sqrt{d}\big\}$ forms an orthonormal basis of observables. 
\en

\bprf
The calculation 
\[
\Tr\big[L_{j}^{2}\big]
=
\Tr\big[4P_{j}^2-4P_{j}+\mathds{1}_{d}\big]
=
\Tr[4P_{j}-4P_{j}+\mathds{1}_{d}]
=d
\]
proves that the $L_{j}/\sqrt{d}$ are normalized for all $j\in\{1,\dots,d^2\}$. 
Now let $j,k\in\{1,\dots,d^2\}$ be distinct. A short calculation shows that 
\[
\Tr[L_{j}L_{k}]
=4\Tr[P_{j}P_{k}]-2\Tr[P_{j}]-2\Tr[P_{k}]+\Tr[\mathds{1}_{d}]=
\frac{4}{d+1}-4+d =d\left(\frac{d-3}{d+1}\right) ,
\]
which is zero precisely when $d=3$, as needed. 
\eprf

Since uncountably many SIC-POVMs exist in dimension 3~\cite{RBKSC04}, Proposition~\ref{prop:LTOSICPOVM} allows us to construct a plethora of examples of orthonormal bases of light-touch observables in dimension 3. The remainder of this paper constructs families of such light-touch observables as well as a generalized pseudo-density matrix, the canonical state over time. Following Ref.~\cite{RBKSC04}, define
\[
\omega:=e^{\frac{2\pi i}{d}}
\qquad\text{ and }\qquad
G_{jk}:=\omega^{\frac{jk}{2}}\sum_{l=0}^{d-1}\omega^{jl}|k\oplus l\>\<l|\, ,
\]
where $\oplus$ denotes addition modulo $d$. Ref.~\cite{RBKSC04} then provides the following characterization of SIC-POVMs in dimension $3$. 

\bn
\label{prop:SICPOVMdim3}
In what follows, given $r_0\in\R$ satisfying $3 r_0^2<2$, set 
\[
r_{\pm}:=\frac{1}{2}\left(r_{0}\pm\sqrt{2-3r_{0}^2}\right).
\]
Let $V,W,F\subset\C^{3}$ denote the sets of vectors
\[
\begin{split}
V&=\left\{\begin{pmatrix}r_{0}\\r_{+}e^{i\theta}\\r_{-}e^{i\varphi}\end{pmatrix}\,:\,\theta,\varphi\in\left\{\frac{\pi}{3},\pi,\frac{5\pi}{3}\right\},\, \frac{1}{\sqrt{2}}<r_{0}\le\sqrt{\frac{2}{3}}\right\},\\
W&=\left\{\frac{1}{\sqrt{2}}\begin{pmatrix}1\\e^{i\chi}\\0\end{pmatrix}\;:\;0\le\chi<2\pi\right\}, \;\text{ and }\;\\
F&=\bigcup_{\sigma\in S_{3}}\big((P_{\sigma}V)\cup (P_{\sigma}W)\big)\, ,
\end{split}
\]
where $P_{\sigma}$ is the permutation matrix associated with $\sigma\in S_{3}$, where $S_{3}$ is the symmetric group on 3 elements. Given any element $|\psi\>\in F$, the set of rank-$1$ projectors 
$\big\{P_{jk}:=G_{jk}|\psi\>\<\psi|G_{jk}^{\dag}\big\}$, with 
$j,k\in\{0,\dots,d-1\}$, 
defines a SIC-POVM. 
\en

We will content ourselves with one example by starting with a fiducial vector $|\psi\>$ in the set $W$ from Proposition~\ref{prop:SICPOVMdim3} with $\chi=0$. This example leads to a SIC-POVM with an experimental implementation~\cites{MTRSTFS11,PMMVDSP13,Tabia12}. For completeness, we write out the (\emph{Weyl--Heisenberg}) matrices $G_{jk}$, the associated unit vectors $|\psi_{jk}\>:=G_{jk}|\psi\>$, and their corresponding light-touch observables $L_{jk}:=2|\psi_{jk}\>\<\psi_{jk}|-\mathds{1}_{d}$ in Table~\ref{table:qutritmatrices}. 
\begin{table*}%[h]
\begin{align*}
G_{00}&=\begin{pmatrix}1&0&0\\0&1&0\\0&0&1\end{pmatrix}
& 
G_{01}&=\begin{pmatrix}0&0&1\\1&0&0\\0&1&0\end{pmatrix}
&
G_{02}&=\begin{pmatrix}0&1&0\\0&0&1\\1&0&0\end{pmatrix}
\\
G_{10}&=\begin{pmatrix}1&0&0\\0&e^{\frac{2\pi i}{3}}&0\\0&0&e^{-\frac{2\pi i}{3}}\end{pmatrix}
&
G_{11}&=\begin{pmatrix}0&0&e^{-\frac{\pi i}{3}}\\e^{\frac{\pi i}{3}}&0&0\\0&-1&0\end{pmatrix}
&
G_{12}&=\begin{pmatrix}0&e^{-\frac{2\pi i}{3}}&0\\0&0&1\\e^{\frac{2\pi i}{3}}&0&0\end{pmatrix}
\\
G_{20}&=\begin{pmatrix}1&0&0\\0&e^{-\frac{2\pi i}{3}}&0\\0&0&e^{\frac{2\pi i}{3}}\end{pmatrix}
&
G_{21}&=\begin{pmatrix}0&0&e^{-\frac{2\pi i}{3}}\\e^{\frac{2\pi i}{3}}&0&0\\0&1&0\end{pmatrix}
&
G_{22}&=\begin{pmatrix}0&e^{\frac{2\pi i}{3}}&0\\0&0&1\\e^{-\frac{2\pi i}{3}}&0&0\end{pmatrix},
\end{align*}
\begin{align*}
|\psi_{00}\>&=\frac{1}{\sqrt{2}}\begin{pmatrix}1\\1\\0\end{pmatrix}
&
|\psi_{01}\>&=\frac{1}{\sqrt{2}}\begin{pmatrix}0\\1\\1\end{pmatrix}
&
|\psi_{02}\>&=\frac{1}{\sqrt{2}}\begin{pmatrix}1\\0\\1\end{pmatrix}
\\
|\psi_{10}\>&=\frac{1}{\sqrt{2}}\begin{pmatrix}1\\e^{\frac{2\pi i}{3}}\\0\end{pmatrix}
&
|\psi_{11}\>&=\frac{1}{\sqrt{2}}\begin{pmatrix}0\\e^{\frac{\pi i}{3}}\\-1\end{pmatrix}
&
|\psi_{12}\>&=\frac{1}{\sqrt{2}}\begin{pmatrix}e^{-\frac{2\pi i}{3}}\\0\\e^{\frac{2\pi i}{3}}\end{pmatrix}
\\
|\psi_{20}\>&=\frac{1}{\sqrt{2}}\begin{pmatrix}1\\e^{-\frac{2\pi i}{3}}\\0\end{pmatrix}
&
|\psi_{21}\>&=\frac{1}{\sqrt{2}}\begin{pmatrix}0\\e^{\frac{2 \pi i}{3}}\\1\end{pmatrix}
&
|\psi_{22}\>&=\frac{1}{\sqrt{2}}\begin{pmatrix}e^{\frac{2\pi i}{3}}\\0\\e^{-\frac{2\pi i}{3}}\end{pmatrix},
\end{align*}
\begin{align*}
L_{00}&=\begin{pmatrix}0&1&0\\1&0&0\\0&0&-1\end{pmatrix}
&
L_{01}&=\begin{pmatrix}-1&0&0\\0&0&1\\0&1&0\end{pmatrix}
&
L_{02}&=\begin{pmatrix}0&0&1\\0&-1&0\\1&0&0\end{pmatrix}
\\
L_{10}&=\begin{pmatrix}0&e^{-\frac{2\pi i}{3}}&0\\e^{\frac{2\pi i}{3}}&0&0\\0&0&-1\end{pmatrix}
&
L_{11}&=\begin{pmatrix}-1&0&0\\0&0&e^{-\frac{2\pi i}{3}}\\0&e^{\frac{2 \pi i}{3}}&0\end{pmatrix}
&
L_{12}&=\begin{pmatrix}0&0&e^{\frac{2 \pi i}{3}}\\0&-1&0\\e^{-\frac{2\pi i}{3}}&0&0\end{pmatrix}
\\
L_{20}&=\begin{pmatrix}0&e^{\frac{2 \pi i}{3}}&0\\e^{-\frac{2\pi i}{3}}&0&0\\0&0&-1\end{pmatrix}
&
L_{21}&=\begin{pmatrix}-1&0&0\\0&0&e^{\frac{2\pi i}{3}}\\0&e^{-\frac{2\pi i}{3}}&0\end{pmatrix}
&
L_{22}&=\begin{pmatrix}0&0&e^{-\frac{2\pi i}{3}}\\0&-1&0\\e^{\frac{2\pi i}{3}}&0&0\end{pmatrix}.
\end{align*}
\caption{The \emph{Weyl--Heisenberg} matrices $G_{jk}$, the associated unit vectors $|\psi_{jk}\>:=G_{jk}|\psi\>$, and their corresponding light-touch observables $L_{jk}:=2|\psi_{jk}\>\<\psi_{jk}|-\mathds{1}_{3}$ for qutrits.}
\label{table:qutritmatrices}
\end{table*}
This set of $9$ matrices $\{L_{jk}\}$ forms an orthogonal basis of light-touch observables in dimension 3 (orthonormal if each is multiplied by $1/\sqrt{3}$). 

Now, given any qutrit density matrix $\rho\in\mathcal{S}(A)$ and any CPTP map $\map{E}:\Alg{A}\to\Alg{B}$, where $\Alg{B}=\matr_{3}$, the state over time associated with the process $(\map{E},\rho)$ is given by 
\be
\label{eq:PDMqutrit}
R(\map{E},\rho)=\frac{1}{d^2}\sum_{j,k,m,n=0}^{2}\<L_{jk},L_{mn}\>L_{jk}\otimes L_{mn}\, ,
\ee
where $\<L_{jk},L_{mn}\>$ is the two-time expectation value with respect to $(\map{E},\rho)$. By Proposition~\ref{prop:LTOSICPOVM}, $R(\map{E},\rho)$ uniquely specifies an operator in $\Alg{A}\otimes\Alg{B}$ representing the spatiotemporal correlations of the two-time expectation values. And by Theorem~\ref{MTXS45739}, this operator is also given by the simple closed-form expression 
$R(\map{E},\rho)=\frac{1}{2}\big\{\rho\otimes\mathds{1}_{\!B},\Jamiol[\map{E}]\big\}$. 
As a concrete example
\[
\rho=\begin{pmatrix}1&0&0\\0&0&0\\0&0&0\end{pmatrix}
\quad\text{ and }\quad
\map{E}=\id_{\matr_{3}}
\]
lead to
\[
R(\map{E},\rho)=
\begin{pmatrix}
1&0&0&0&0&0&0&0&0\\
0&0&0&\frac{1}{2}&0&0&0&0&0\\
0&0&0&0&0&0&\frac{1}{2}&0&0\\
0&\frac{1}{2}&0&0&0&0&0&0&0\\
0&0&0&0&0&0&0&0&0\\
0&0&0&0&0&0&0&0&0\\
0&0&\frac{1}{2}&0&0&0&0&0&0\\
0&0&0&0&0&0&0&0&0\\
0&0&0&0&0&0&0&0&0\\
\end{pmatrix}\, ,
\]
which has eigenvalues $\{1,-1/2,-1/2,1/2,1/2,0,0,0,0\}$ (compare with the qubit case in (7) of Ref.~\cite{FJV15}).

Meanwhile, Equation~\eqref{eq:PDMqutrit} involves the computation of $(d^{2})^{2}=d^{4}=3^4=81$ terms, and is useful when the two-time expectation values are known, say, from experimental data. On the other hand, when both $\rho$ and $\map{E}$ are known, the computation of $R(\map{E},\rho)$ is achieved almost immediately using the expression $R(\map{E},\rho)=\frac{1}{2}\big\{\rho\otimes\mathds{1}_{\!B},\Jamiol[\map{E}]\big\}$, thus illustrating its utility. Therefore, we hope that our Theorem~\ref{MTXS45739} extending the pseudo-density matrix formalism to arbitrary dimensions, together with its simple expression through the usage of quantum states over time, may expand our understanding of spatiotemporal correlations for quantum systems beyond multiqubit systems. 

\begin{figure*}[t]
%\begin{subfigure}{0.99\textwidth}
%\centering
\includegraphics[width=\textwidth]{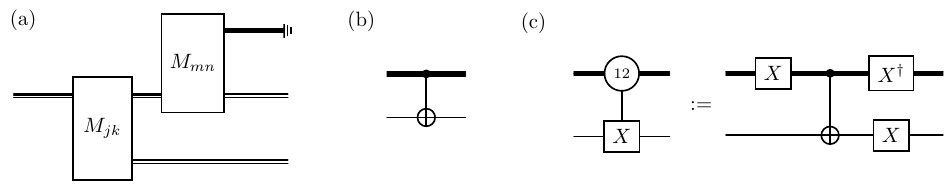}
%
%\end{subfigure}
%\begin{subfigure}{0.99\textwidth}
%\centering
\includegraphics[width=\textwidth]{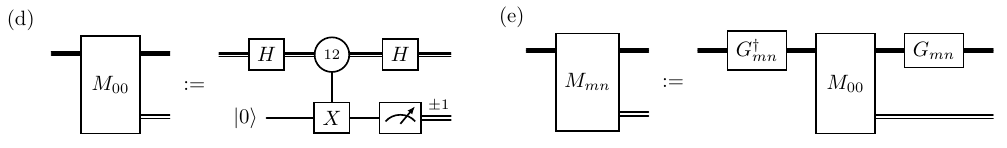}
%\end{subfigure}
\caption{
\textbf{(a)} The general sequential measurement protocol where Alice measures $L_{jk}$ on a qutrit system, records outcome $\pm 1$ on a classical register, and sends the updated state to Bob, who then measures $L_{mn}$ and records outcome $\pm 1$ on a classical register (the $M_{jk}$ are defined in (e)).
\textbf{(b)} A qubit-qutrit hybrid \emph{controlled-sum} gate sending $|j\>\otimes|k\>$ to $|j+k\mod 2\>\otimes|k\>$~\cites{GoPr04,SZBY22,CDBSCC23}.
\textbf{(c)} A modified \emph{controlled-$X$} gate $\mathds{1}_{2}\otimes|0\>\<0|+X\otimes\big(|1\>\<1|+|2\>\<2|\big)$, where the $12$ indicates the basis states for which $X$ is applied. 
The gate $X$ on the top (qutrit) wire is the \emph{bit-shift} $X=G_{01}$ defined as $X|j\>=|j+1\mod3\>$. The unitary $X$ gate on the bottom (qubit) wire on the right is the standard Pauli gate $X|j\>=|j+1\mod2\>$. 
\textbf{(d)} An implementation $M_{00}$ of the measurement of $L_{00}$ on a qutrit using an ancillary qubit set initially to $|0\>$, the latter of which is then measured in the computational basis. The gate $H$ on the top wire is the \emph{ternary extension} of the Hadamard gate and is given by $H|0\>=\frac{1}{\sqrt{2}}\big(|0\>+|1\>\big)$, $H|1\>=\frac{1}{\sqrt{2}}\big(|0\>-|1\>\big)$, and $H|2\>=|2\>$. 
\textbf{(e)} An implementation $M_{mn}$ of the measurement of $L_{mn}$ on a qutrit expressed in terms of the measurement $M_{00}$ and the unitary gates $G_{mn}$, which are defined after Proposition~\ref{prop:SICPOVMdim3}. Note that $G_{mn}=X^{n}Z^{m}$ for all $m,n$ (up to an overall phase), where $Z$ is the qutrit \emph{phase} gate $Z|j\>=e^{2\pi ij/3}|j\>$~\cite{MTRSTFS11}.
}
\label{fig:seqmeasLT}
\end{figure*}

Before concluding this section, we briefly propose quantum circuit implementations for measuring the two-time expectation values for qutrits in Figure~\ref{fig:seqmeasLT}. In this figure, we follow the convention used in Qiskit~\cite{qiskit2024}, in that quantum circuits are read from left to right in time and from bottom to top using the tensor product. 
In all diagrams, the top (thick) solid wire is a qutrit, the bottom (thin) solid wire is a qubit, and double-wires are classical registers. 
The different gates appearing in Figure~\ref{fig:seqmeasLT} can be obtained from elementary qutrit gates~\cites{DiWe12,RLKS23,BPDM23,LanyonSimplifying2009,LTCL23,SZBY22,MuSt00}. 
For a single measurement of $L_{jk}$, one could use a single qutrit system and one classical register to record the output, or one could use a single qutrit system, an ancillary qubit, and a classical register, where the measurement is performed on the ancillary qubit instead, since the measurement is, after all, $\pm 1$-valued. In the circuits in Figure~\ref{fig:seqmeasLT}, the latter has been used.

\section{Discussion}

In this work, we have extended the pseudo-density matrix formalism of Ref.~\cite{FJV15} to arbitrary finite-dimensional systems by making use of the construct of a canonical state over time. While the canonical state over time has been derived from physically motivated axioms as first introduced in Ref.~\cite{HHPBS17}, the present work solidifies the operational meaning of this operator as encoding the two-time expectation values of light-touch observables (cf. Theorem~\ref{MTXS45739} and Remark~\ref{OPXINTXS347}), which one may view as a generalization of Pauli observables to arbitrary finite-dimensional systems. We also showed that light-touch observables are informationally complete in the sense that there exists a basis consisting of light-touch observables for every finite-dimensional quantum system (cf. Proposition~\ref{prop:LTIC}), and we constructed an explicit orthonormal basis of light-touch observables for qutrit systems in terms of SIC-POVMs.

During the course of this work, the operator representation in Theorem~\ref{MTXS45739} has been utilized in several different ways. For example, Ref.~\cite{SongParzygnat2025} makes use of Theorem~\ref{MTXS45739} to derive an explicit test to determine whether or not the correlations between local measurements performed on systems $A$ and $B$ admit a causal explanation. 
Namely, given data consisting of the \emph{products} of local measurements of light-touch observables (based on many runs of some experiment), the resulting averages uniquely determine an operator by the informational completeness of light-touch observables in a way that is analogous to quantum state tomography. Ref.~\cite{SongParzygnat2025} then derives necessary and sufficient conditions for this operator to coincide with the canonical state over time, which in turn provides necessary and sufficient conditions for such data to be compatible with direct causal influence due to Theorem~\ref{MTXS45739}.
This result gives a potential quantum advantage over the analogous question in classical causal inference, since there are cases where a quantum correlation could distinguish between mere correlation and actual causation. As a second example, note that the linear functional on $\Alg{A}\otimes\Alg{B}$ uniquely determined by $X_{\!A}\otimes X_{\!B}\mapsto \<\!\<X_{\!A},X_{\!B}\>\!\>:={\Tr\big[(\mathcal{E}\star\rho)(X_{\!A}\otimes X_{\!B})\big]}$ coincides with the real part of the standard two-point correlator $\Tr\big[\rho X_{\!A}\mathcal{E}^*(X_{\!B})\big]$ as first introduced by Dirac~\cite{Di45}. Such a correlator has its own operational interpretation as described in Refs.~\cites{BDOV13,BDOV14}, and also specializes to the \emph{Margenau--Hill} quasi-probability distribution when $X_A$ and $X_B$ are taken as projection operators~\cite{FuMaWu2025}. Our work shows that although the correlator $\<\!\<\obs{A},\obs{B}\>\!\>$ is not equal to the two-time expectation value $\<\obs{A},\obs{B}\>$ for general observables $\obs{A}\in\mathbf{Obs}(A)$ and $\obs{B}\in\mathbf{Obs}(B)$, the two quantities agree whenever $\obs{A}$ is a light-touch observable. 
Besides the above applications, the state over time $\mathcal{E}\star\rho$ has also been used in quantum optimal transport as a dynamical quantum coupling~\cite{HoCaWi25}, whereas earlier approaches utilized canonical purifications~\cite{PaTr21} (also called canonical compound states~\cites{ChMa20,MaCh22}).

This work opens up several avenues for further exploration. 
(1) The first natural question concerns the extension of our results to infinite-dimensional quantum systems. 
(2) A second question is if there exist other bases of observables for which an analogous $R(\map{E},\rho)$ can be constructed and defines a quantum state over time satisfying a subset of the axioms from Refs.~\cites{FuPa22,FuPa22a}. 
Such a state over time cannot, in general, be given by the expression ${\frac{1}{2}\big\{\rho\otimes\mathds{1}_{\!B},\Jamiol[\map{E}]\big\}}$, and therefore it cannot satisfy the characterizing properties of Refs.~\cites{LiNg23,PFBC23}. 
Nevertheless, for the purposes of allowing measurements other than light-touch observables, it is important to explore such possibilities for analyzing spatiotemporal correlations. 

\vspace{3mm}
\noindent
{\bf Data Accessibility.}
No new data were created or analyzed in this study. Data sharing is not applicable to this article.

\vspace{3mm}
\noindent
{\bf Conflict of interests.}
None.

\vspace{3mm}
\noindent
{\bf Funding.}
JF is supported by the Key Development Project of Hainan Province for the project ``Spacetime from Quantum Information", grant no. 126MS0010.

\vspace{3mm}
\noindent
{\bf Acknowledgements.}
We thank Matthew Pusey and Minjeong Song for discussions. 

%%%%%%%%BIBLIOGRAPHY%%%%%%%%%%%%
\addcontentsline{toc}{section}{\numberline{}Bibliography}
 \bibliographystyle{plain}
\bibliography{OTX}

\end{document}